\documentclass[aps,rpl,preprint,groupedaddress]{revtex4}
\begin{document}

\title{Teleportation in the presence of noise}
\author{Ye Yeo}
\affiliation{Department of Physics, National University of Singapore, 10 Kent Ridge Crescent, Singapore 119260, Singapore}

\author{Zhe-Wei Kho}
\affiliation{Hwa Chong Institution, 661 Bukit Timah Road, Singapore 269734, Singapore}

\author{Lixian Wang}
\affiliation{Hwa Chong Institution, 661 Bukit Timah Road, Singapore 269734, Singapore}

\begin{abstract}
Non-commuting noises may give rise to entanglement sudden death.  By considering the decoherence dynamics during establishment of the channel states 
and noisy recovery operations, we study further the impact of non-commuting noises on single- and two-qubit teleportation.  We show that in the 
presence of these noises there exists a critical rate of recovery operation below which teleportation will fail.
\end{abstract}

\maketitle

Quantum teleportation \cite{Bennett} is the disembodied transport of an arbitrary unknown state $|\psi\rangle$ of a quantum system from a sender 
(Alice, $A$) to a receiver (Bob, $B$) using their prior shared entangled state $\chi$.  We focus on two-level systems or qubits throughout this 
work.  So,
\begin{equation}
|\psi\rangle = a|0\rangle + b|1\rangle
\end{equation}
with $a = \cos\theta/2$ and $b = e^{i\phi}\sin\theta/2$, where $0 \leq \theta \leq \pi$ and $0 \leq \phi \leq 2\pi$.  In the standard teleportation 
protocol, Alice first performs a joint measurement $\{\Pi^j \equiv |\Psi^j_{\rm Bell}\rangle\langle\Psi^j_{\rm Bell}|\}$ in the Bell operator basis:
\begin{eqnarray}
|\Psi^0_{\rm Bell}\rangle & = & \frac{1}{\sqrt{2}}(|00\rangle + |11\rangle), \nonumber \\
|\Psi^1_{\rm Bell}\rangle & = & \frac{1}{\sqrt{2}}(|01\rangle + |10\rangle), \nonumber \\
|\Psi^2_{\rm Bell}\rangle & = & \frac{1}{\sqrt{2}}(|01\rangle - |10\rangle), \nonumber \\
|\Psi^3_{\rm Bell}\rangle & = & \frac{1}{\sqrt{2}}(|00\rangle - |11\rangle),
\end{eqnarray}
on particle $A_1$ in the state $|\psi\rangle$ and particle $A_2$ in the entangled channel state $\chi$.  She then communicates the measurement outcome 
$j$ $(j = 0, 1, 2, 3)$ to Bob via a classical channel.  Depending on Alice's measurement outcome, Bob either does nothing or applies a recovery 
operation to particle $B$ also in the state $\chi$ to complete the teleportation.  Ideally, these operations consist of $\sigma^1$, $\sigma^2$ 
and $\sigma^3$, the usual Pauli operators that satisfy the commutation relations $[\sigma^m, \sigma^n] = 2i\epsilon^{lmn}\sigma^l$.  Teleportation is 
central to a number of fundamental quantum communication and computation schemes.  However, for a large-scale realization of these, it is necessary to 
be able to teleport more than one qubit.  The standard teleportation protocol may be directly generalized to teleport an $N$-qubit state via $N$ copies 
of $\chi$ \cite{Rigolin}.  The first experimental realization of teleportation of two-qubit system has been presented in Ref.\cite{Zhang}.  In 
this paper, we consider the dynamics of single- and two-qubit teleportation in the presence of noise.

Real systems can never be perfectly isolated from the surrounding world.  Therefore, in a realistic study of any quantum information protocol, one has 
to take into account the unavoidable coupling of systems or quantum processors with their environment.  Under the assumption that the environment is 
Markovian, the state $\rho$ of an open system evolves according to a quantum master equation \cite{Breuer}
\begin{equation}\label{master}
\frac{d}{dt}\rho = -i[H, \rho] + \sum_k\frac{1}{2}\gamma_k(2L_k\rho L^{\dagger}_k - L^{\dagger}_kL_k\rho - \rho L^{\dagger}_kL_k).
\end{equation}
Here, we set $\hbar = 1$.  $H$ is the system's Hamiltonian.  In particular, $H = H_m \equiv \omega_0\sigma^m/2$ generates an anticlockwise coherent 
rotation of a qubit about the $m$-axis at the rate $\omega_0$.  This may be a necessary unitary transformation for Bob to recover $|\psi\rangle$.  
$L_k$'s are the Lindblad operators.  They describe decoherence.  For instance, consider a pair of qubits initially in the maximally entangled Bell 
singlet state $|\Psi^2_{\rm Bell}\rangle$.  If one of the particles is subject to a local dephasing noise, say $L_{03} = \sigma^0 \otimes \sigma^3$ 
($\sigma^0 = I_2$ is the two-dimensional identity), then solving the master equation, we find $\zeta(t)$ at time $t > 0$:
\begin{equation}\label{diag0}
|\Psi^2_{\rm Bell}\rangle\langle\Psi^2_{\rm Bell}| \rightarrow \zeta(t) = 
\frac{1}{2}\left(\begin{array}{cccc}
0 & 0 & 0 & 0 \\
0 & 1 & -e^{-2\gamma_3 t} & 0 \\
0 & -e^{-2\gamma_3 t} & 1 & 0 \\
0 & 0 & 0 & 0
\end{array}\right).
\end{equation}
When one of the particles is instead subject to a local bit-flip noise, i.e., $L_{01} = \sigma^0 \otimes \sigma^1$, then
\begin{equation}\label{diagnot0}
|\Psi^2_{\rm Bell}\rangle\langle\Psi^2_{\rm Bell}| \rightarrow \xi(t) = 
\frac{1}{4}\left(\begin{array}{cccc}
1 - e^{-2\gamma_1 t} & 0 & 0 & -(1 - e^{-2\gamma_1 t}) \\
0 & 1 + e^{-2\gamma_1 t} & -(1 + e^{-2\gamma_1 t}) & 0 \\
0 & -(1 + e^{-2\gamma_1 t}) & 1 + e^{-2\gamma_1 t} & 0 \\
-(1 - e^{-2\gamma_1 t}) & 0 & 0 & 1 - e^{-2\gamma_1 t}
\end{array}\right).
\end{equation}
Equations (\ref{diag0}) and (\ref{diagnot0}) describe the physical situation where Alice prepares, at time $t = 0$, the state 
$|\Psi^2_{\rm Bell}\rangle$ and sends one of the particles down a noisy channel to Bob to establish the shared entangled state $\chi$ at $t > 0$.  We 
say the mixed channel state $\chi$ results from transmission noise.

For a two-qubit system $AB$, it is well-known that a necessary and sufficient condition for separability is that a matrix, obtained by partial 
transposition of its density operator, has non-negative eigenvalue(s) \cite{Peres}.  As a measure of the amount of entanglement associated with a 
given two-qubit state $\rho_{AB}$, we employ the negativity \cite{Vidal}
\begin{equation}
{\cal N}[\rho_{AB}] \equiv \max\left\{0, -2\sum_m\lambda_m\right\},
\end{equation}
where $\lambda_m$ is a negative eigenvalue of $\rho^{T_A}_{AB}$, the partial transposition of $\rho_{AB}$.  From Eqs.(\ref{diag0}) and (\ref{diagnot0}), 
we obtain ${\cal N}[\zeta(t)] = e^{-2\gamma_3t}$ and ${\cal N}[\xi(t)] = e^{-2\gamma_1t}$.  The entanglement associated with 
$|\Psi^2_{\rm Bell}\rangle$ decays smoothly and asymptotically exponentially to zero, in exactly the same manner whether it is under the influence of a 
dephasing or bit-flip noise.  This can be understood from the facts that $|0\rangle$ and $|1\rangle$ are eigenvectors of $\sigma^3$, and that $|\Psi^2_{\rm Bell}\rangle$ can also be 
similarly expressed in terms of the eigenvectors $|\pm\rangle \equiv (|0\rangle \pm |1\rangle)/\sqrt{2}$ of $\sigma^1$:
\begin{equation}
|\Psi^2_{\rm Bell}\rangle = \frac{1}{\sqrt{2}}(|-\rangle|+\rangle - |+\rangle|-\rangle).
\end{equation}
But, if we subject $|\Psi^2_{\rm Bell}\rangle$ simultaneously to both $L_{01}$ and $L_{03}$, then we find a dramatically different effect.  The 
resulting state $\mu(t)$ has negativity \cite{remark1}
\begin{equation}
{\cal N}[\mu(t)] = \max\left\{0, \frac{1}{2}[(1 + e^{-2\gamma t})^2 - 2]\right\}.
\end{equation}
From here on, we let $\gamma_k = \gamma$ for simplicity.  The entanglement goes abruptly to zero in a finite time 
$\tau_{\rm d} = -\ln(\sqrt{2} - 1)/(2\gamma)$ and remains zero thereafter.  This phenomenon, called ``entanglement sudden death" (ESD) \cite{Yu}, has 
recently been confirmed experimentally \cite{Eberly, Almeida}.  To observe ESD under dephasing, an entangled two-qubit density operator should have 
nonzero diagonal elements \cite{Huang}.  It is thus not difficult to understand how ESD occurs here.  Since $\sigma^1$ does not commute with $\sigma^3$, 
the noise generated by $\sigma^1$ causes zero diagonal elements like those in Eq.(\ref{diag0}) to become nonzero like those in Eq.(\ref{diagnot0}).  
Since they do not commute, we call the noises generated by $\sigma^1$ and $\sigma^3$ non-commuting.  These have potentially troubling consequence 
\cite{Eberly}.

Given the fundamental and practical importance of teleportation, it is imperative to understand all possible environmental effects on the protocol 
(see Refs.\cite{Oh, Carlo, Lee} for some of these studies).  In this paper, motivated by the above experimental progresses and theoretical insights, we 
analyze the impact of non-commuting noises and non-commuting recovery operations in the presence of different types of noise on single- and two-qubit 
teleportation.  We show that in the presence of both transmission and recovery noises, there exists a critical $\omega_0$ below which the teleportation 
will fail.  For two-qubit teleportation, we show that even when the channel states are ideal, the entanglement associated with a certain class of 
entangled input states suffers from ESD whenever Bob's noisy recovery operations do not commute.

To set the stage and for definiteness, we suppose the ideal channel state is $\chi_0 \equiv |\Psi^2_{\rm Bell}\rangle\langle\Psi^2_{\rm Bell}|$.  Then, 
the initial complete state of the three particles, $A_1$, $A_2$ and $B$, is
\begin{equation}\label{teleport}
|\psi\rangle \otimes |\Psi^2_{\rm Bell}\rangle = \frac{1}{2}(
-i|\Psi^0_{\rm Bell}\rangle \otimes \sigma^2|\psi\rangle - |\Psi^1_{\rm Bell}\rangle \otimes \sigma^3|\psi\rangle 
- |\Psi^2_{\rm Bell}\rangle \otimes \sigma^0|\psi\rangle + |\Psi^3_{\rm Bell}\rangle \otimes \sigma^1|\psi\rangle).
\end{equation}
It follows ideally that if Alice's measurement outcome is $j$, then Bob's unitary transformation to recover $|\psi\rangle$ will be 
$\sigma^m$ with $m = j \oplus 2$.  ($\oplus$ is addition modulo four.)  We assume throughout that only Alice's joint Bell-state measurement is perfect.  
In general, the unnormalized state teleported via a noisy channel state $\chi$ is therefore given by
\begin{equation}\label{single}
{\cal E}^{(\alpha)}_{\chi, m}(|\psi\rangle\langle\psi|) = {\cal R}^{(\alpha)}_m({\rm tr}_{A_1A_2}[(\Pi^j_{A_1A_2} \otimes \sigma^0_B)
(|\psi\rangle_{A_1}\langle\psi| \otimes \chi_{A_2B})]).
\end{equation}
$\alpha = p$, $d$, $b$, $bp$ or $i$ indicates if Bob's recovery operations are {\it p}erfect, or corrupted by {\it d}ephasing, {\it b}it-flip, 
{\it b}it-{\it p}hase-flip noise (generated by $\sigma^2$), or {\it i}ntrinsic noise.  For $\alpha = p$, ${\cal R}^{(p)}_m(\rho) = 
\sigma^m\rho\sigma^m$.  Otherwise, ${\cal R}^{(\alpha)}_m(\rho)$ has to be obtained by solving the appropriate master equation (\ref{master}) with 
$\rho$ as the initial state.  By intrinsic noise we refer to Milburn's model of intrinsic decoherence \cite{Milburn}, where the Lindblad operator in 
Eq.(\ref{master}) is given by the Hamiltonian $H$.

Quantitatively, the average teleportation fidelity
\begin{equation}\label{fidelity}
{\cal F}^{(\alpha)}_{\rm av}[\chi] \equiv \frac{1}{4\pi}\int^{\pi}_0\sin\theta d\theta\int^{2\pi}_0d\phi\ 
                                   \sum^3_{m = 0}\langle\psi|{\cal E}^{(\alpha)}_{\chi, m}(|\psi\rangle\langle\psi|)|\psi\rangle
\end{equation}
describes if the teleportation protocol is successful.  For example, when $\chi = \zeta(t)$ or $\xi(t)$ but $\alpha = p$, we have 
${\cal F}^{(p)}_{\rm av}[\zeta(t)] = (2 + e^{-2\gamma t})/3 = {\cal F}^{(p)}_{\rm av}[\xi(t)]$ \cite{Oh}.  In both cases, 
${\cal F}^{(p)}_{\rm av}[\chi]$ decays smoothly and asymptotically exponentially to the limiting value of $2/3$ - the best possible score if Alice and 
Bob communicate with each other only via a classical channel.  In contrast, we have for $\chi = \mu(t)$,
\begin{equation}\label{res0}
{\cal F}^{(p)}_{\rm av}[\mu(t)] = \frac{2}{3} + \frac{1}{6}[(1 + e^{-2\gamma t})^2 - 2].
\end{equation}
We note that ${\cal F}^{(p)}_{\rm av}[\mu(t)] = 2/3$ when $t = \tau_{\rm d}$, exactly at the moment when the entanglement of $\mu$ suffers 
a sudden death.  Entanglement is a necessary resource for teleportation, and provided Bob's recovery operations are perfect, every bit of 
entanglement associated with $\zeta(t)$, $\xi(t)$, or $\mu(t)$ will yield an average teleportation fidelity better than classical communication alone 
does.  The protocol can still be successfully completed with Bob's perfect recovery operations if the channel state $\mu $ is established within 
$\tau_{\rm d}$.  ESD sets a limit on the success of the protocol, in terms of the finite lifetime $\tau_{\rm d}$ of the entanglement 
associated with the channel state $\mu$.  

Before analyzing the effects of both transmission and recovery noises, let us suppose the transmission is noiseless and Alice shares $\chi_0$ with Bob,
while Bob's recovery operations are noisy.  Firstly, if Bob's operations are corrupted by intrinsic noise \cite{Milburn},  then we have
\begin{equation}\label{res1}
{\cal F}^{(i)}_{\rm av}[\chi_0] = \frac{1}{4}(3 - e^{-2\gamma t}\cos\omega_0t).
\end{equation}
The maxima of ${\cal F}^{(i)}_{\rm av}[\chi_0]$ are reached whenever $t$ satisfies $2\gamma\cos\omega_0t + \omega_0\sin\omega_0t = 0$ and 
$(\omega^2_0 - 4\gamma^2)\cos\omega_0t - 4\omega_0\gamma\sin\omega_0t < 0$.  The first (and largest) maximum, $\Phi^{(i)}_{\max}[\chi_0]$, is achieved 
at $T^{(i)}_{\chi_0} = (\pi - \tan^{-1}2\gamma/\omega_0)/\omega_0$.  For a given $\gamma$, $T^{(i)}_{\chi_0}$ increases with decreasing $\omega_0$.  
However, $\Phi^{(i)}_{\max}[\chi_0]$ is always greater than $2/3$ by at least $1/12$.  Secondly, if Bob's operations are instead infected with 
dephasing noise, then
\begin{equation}\label{res2}
{\cal F}^{(d)}_{\rm av}[\chi_0] = \frac{1}{12}[(8 + e^{-2\gamma t}) - e^{-2\gamma t}\cos\omega_0t - 2e^{-\gamma t}\cos\omega t],
\end{equation}
where $\omega \equiv \sqrt{\omega^2_0 - \gamma^2}$.  In this case, the maxima of ${\cal F}^{(d)}_{\rm av}[\chi_0]$ are reached when $t$ satisfies 
$e^{-\gamma t}[(2\gamma\cos\omega_0t + \omega_0\sin\omega_0t) - 2\gamma] + 2(\gamma\cos\omega t + \omega\sin\omega t) = 0$ and 
$e^{-\gamma t}[(\omega^2_0 - 4\gamma^2)\cos\omega_0t - 4\omega_0\gamma\sin\omega_0t + 4\gamma^2] 
+ 2[(\omega^2 - \gamma^2)\cos\omega t - 2\omega\gamma\sin\omega t] < 0$.  Again, the first maximum $\Phi^{(d)}_{\max}[\chi_0]$ is achieved at 
$T^{(d)}_{\chi_0}$ that increases with decreasing $\omega_0$.  But, in contrast to Eq.(\ref{res1}), $\Phi^{(d)}_{\max}[\chi_0]$ decays smoothly and 
asymptotically to $2/3$ with increasing $T^{(d)}_{\chi_0}$.  Hence, for both instances, provided Alice and Bob share an ideal channel state, Bob's noisy 
operations do not result in an average teleportation fidelity worse than that achievable by classical communication alone.  Specifically, there is no 
constraint on the rate at which Bob's operations have to be completed before the teleportation fails.  The difference between Eqs.(\ref{res1}) and 
(\ref{res2}) can be accounted for by the facts that in Milburn's model of decoherence, the Lindblad operators $L_m = \sigma^m$ commute with the 
Hamiltonian $H_m$ by definition; whereas in the presence of dephasing noise, except for $H_3$, $H_1$ and $H_2$ do not commute with the Lindblad operator 
$L_3$.  Finally, we note that ${\cal F}^{(b)}_{\rm av}[\chi_0] = {\cal F}^{(d)}_{\rm av}[\chi_0]$.

Now, we consider $\chi = \zeta(t_0)$ with Bob's operations executed in the presence of intrinsic noise.  After some straightforward 
calculations, we obtain
\begin{equation}\label{res3}
{\cal F}^{(i)}_{\rm av}[\zeta(t_0)] = \frac{1}{12}[(7 + 2e^{-2\gamma t_0}) - (1 + 2e^{-2\gamma t_0})e^{-2\gamma t}\cos\omega_0t].
\end{equation}
Clearly, the first maximum $\Phi^{(i)}_{\max}[\zeta(t_0)]$ is achieved at $T^{(i)}_{\zeta(t_0)} = T^{(i)}_{\chi_0}$.  However, in contrast to 
Eq.(\ref{res1}), we have to demand that
\begin{equation}\label{con1}
(1 + 2e^{-2\gamma t_0})\left\{1 - \exp\left[-2\gamma T^{(i)}_{\zeta(t_0)}\right]\cos\omega_0T^{(i)}_{\zeta(t_0)}\right\} > 2,
\end{equation}
in order for $\Phi^{(i)}_{\max}[\zeta(t_0)]$ to exceed $2/3$.  Equation (\ref{con1}) gives the constraint on a critical $\omega_0$, denoted by 
$\omega^{(i)}_c[\zeta(t_0)]$, smaller than which will result in a failure to teleport.  Suppose $\gamma = 1/10$ and it takes $t_0 = 10$ for Alice's 
entangled qubit to reach Bob so as to establish the entangled channel, then we have to require that $\omega^{(i)}_c[\zeta(t_0)] \approx 1.09915$.  
Therefore, in the presence of both transmission and recovery noises, there exists a critical $\omega_0$ below which the teleportation will fail.  We 
attribute this to the non-commutativity between $L_3$, generator of the transmission noise, and the noisy recovery operations ${\cal R}^{(i)}_1$ and 
${\cal R}^{(i)}_2$.  We note that we could have expressed our results in terms of the transmission time $t_0$ if we fix 
$\omega_0$ instead - there will then exist a critical $t_0$ beyond which the teleportation will fail, just like in Eq.(\ref{res0}).

Next, if Bob's operations are corrupted by dephasing noise,
\begin{eqnarray}\label{res4}
{\cal F}^{(d)}_{\rm av}[\zeta(t_0)] & = & 
\frac{1}{12}[(7 + e^{-2\gamma t_0}) + e^{-2\gamma t_0}e^{-2\gamma t}(1 - \cos\omega_0t) \nonumber \\
& &        - (1 + e^{-2\gamma t_0})e^{-\gamma t}\cos\omega t - (1 - e^{-2\gamma t_0})\frac{\gamma}{\omega}e^{-\gamma t}\sin\omega t].
\end{eqnarray}
In this case, the critical $\omega^{(d)}_c[\zeta(t_0)] \approx 0.754443$ if $\gamma = 1/10$ and $t_0 = 10$.  This is again attributable to the 
fact that ${\cal R}^{(d)}_1$ and ${\cal R}^{(d)}_2$ do not commute with $L_3$.  Lastly, when the operations are infected with bit-flip noise,
\begin{eqnarray}\label{res5}
{\cal F}^{(b)}_{\rm av}[\zeta(t_0)] & = & \frac{1}{24}[(13 + 3e^{-2\gamma t_0}) + (1 + e^{-2\gamma t_0})e^{-2\gamma t}(1 - \cos\omega_0t) \nonumber \\
& & - (1 + 3e^{-2\gamma t_0})e^{-\gamma t}\cos\omega t + (1 - e^{-2\gamma t_0})\frac{\gamma}{\omega}e^{-\gamma t}\sin\omega t].
\end{eqnarray}
We note that ${\cal F}^{(bp)}_{\rm av}[\zeta(t_0)] = {\cal F}^{(b)}_{\rm av}[\zeta(t_0)]$.  For $\gamma = 1/10$ and $t_0 = 10$, we have 
$\omega^{(b)}_c[\zeta(t_0)] \approx 1.38597$.  In contrast to Eqs.(\ref{res3}) and (\ref{res4}), Eq.(\ref{res5}) is due to non-commutativity between 
$L_3$ and all three noisy recovery operations.  Preliminary numerical results indicate that $\omega^{(b)}_c[\zeta(t_0)]$ is larger than both 
$\omega^{(i)}_c[\zeta(t_0)]$ and $\omega^{(d)}_c[\zeta(t_0)]$ regardless of $\gamma$ and $t_0$, while $\omega^{(d)}_c[\zeta(t_0)]$ may be greater than 
$\omega^{(i)}_c[\zeta(t_0)]$ for some $\gamma$ and $t_0$ \cite{remark2}.  We must stress that we have achieved what we set out to show, namely that in 
the presence of both transmission and recovery noises there exists a critical $\omega_0$ below which the teleportation fails to attain an average 
fidelity better than classically possible.

Before concluding, we consider two-qubit teleportation.  A direct generalization of Eq.(\ref{single}) gives
\begin{equation}
{\cal E}^{(\alpha)}_{\chi, mn}(|\Psi\rangle\langle\Psi|) = {\cal R}^{(\alpha)}_{mn}
({\rm tr}_{A_1A_2A_3A_4}[(\Pi^{j_1}_{A_1A_3} \otimes \Pi^{j_2}_{A_2A_4} \otimes I_4)
(|\Psi\rangle_{A_1A_2}\langle\Psi| \otimes \chi_{A_3B_1} \otimes \chi_{A_4B_2})]),
\end{equation}
the unnormalized state teleported via the pair of channel states $\chi \otimes \chi$.  As in Eq.(\ref{single}), for outcomes $j_1$ and $j_2$,
${\cal R}^{(p)}_{mn}(\rho) = (\sigma^m \otimes \sigma^n)\rho(\sigma^m \otimes \sigma^n)$, where $m = j_1 \oplus 2$ and $n = j_2 \oplus 2$.  Otherwise, 
${\cal R}^{(\alpha)}_m(\rho)$ has to be obtained by solving the appropriate master equation (\ref{master}) with $\rho$ as the initial state.  In 
two-qubit teleportation, it is not only important to achieve a high fidelity 
${\cal F}^{(\alpha)}_{mn}[\chi, |\Psi\rangle] \equiv \langle\Psi|\tilde{{\cal E}}^{(\alpha)}_{\chi, mn}(|\Psi\rangle\langle\Psi|)|\Psi\rangle$ 
\cite{remark3}, but also a high negativity 
${\cal N}[\tilde{{\cal E}}^{(\alpha)}_{\chi, mn}(|\Psi\rangle\langle\Psi|)]$ if the input state $|\Psi\rangle$ originally has non-zero entanglement.  
We want to show that to achieve both there also necessarily exists a critical $\omega_0$.  To this end, it is sufficient to consider 
(with $0 \leq \theta \leq \pi$)
\begin{equation}
|\Psi\rangle = \cos\theta|00\rangle + \sin\theta|11\rangle,
\end{equation}
which has negativity ${\cal N}[|\Psi\rangle\langle\Psi|] = |\sin2\theta| \equiv \eta(\theta)$.

Firstly, we consider $\chi = \zeta(t)$ but $\alpha = p$.  In this case, regardless of Alice's measurement outcomes and Bob's corresponding recovery 
operations, we have $\tilde{{\cal E}}^{(p)}_{\zeta(t)}(|\Psi\rangle\langle\Psi|) = 
\cos^2\theta|00\rangle\langle 00| + e^{-4\gamma t}\cos\theta\sin\theta|00\rangle\langle 11| + 
e^{-4\gamma t}\cos\theta\sin\theta|11\rangle\langle 00| + \sin^2\theta|11\rangle\langle 11|$.  It follows that 
${\cal N}[\tilde{{\cal E}}^{(p)}_{\zeta(t)}(|\Psi\rangle\langle\Psi|)] = e^{-4\gamma t}\eta(\theta)$ and 
${\cal F}^{(p)}[\zeta(t), |\Psi\rangle] = 1 - 1/2(1 - e^{-4\gamma t})\eta^2(\theta)$.  So, like in single-qubit teleportation, every bit of 
entanglement associated with $\zeta(t)$ can be used to teleport some entanglement of $|\Psi\rangle$ and with fidelity better than the classically 
achievable $2/5$ \cite{Zhang}.  The same conclusions can be made with $\chi = \xi(t)$.  However, we have for $\chi = \mu(t)$,
\begin{equation}
{\cal N}[\tilde{{\cal E}}^{(p)}_{\mu(t)}(|\Psi\rangle\langle\Psi|)] = \frac{1}{2}[e^{-4\gamma t}(1 + e^{-4\gamma t})\eta(\theta) - (1 - e^{-4\gamma t})].
\end{equation}
$\tilde{{\cal E}}^{(p)}_{\mu(t)}(|\Psi\rangle\langle\Psi|)$ suffers ESD at $\tau'_{\rm d} = -\ln\{[\sqrt{\eta^2 + 6\eta + 1} - (\eta + 1)]/(2\eta)\}/(4\gamma)$.  
$\tau'_{\rm d}$ increases with increasing $\eta$.  In particular, for $\theta = \pi/4$ or $\eta = 1$, $\tau'_{\rm d} = \tau_{\rm d}/2$.  Hence, the 
two-qubit teleportation scheme fails to teleport any entanglement before $\mu$ suffers ESD.  This is in contrast to Eq.(\ref{res0}).  The 
teleportation fidelity at a given time $t$,
\begin{equation}
{\cal F}^{(p)}[\mu(t), |\Psi\rangle] = \frac{1}{4}[(1 + e^{-2\gamma t})^2 - e^{-2\gamma t}(2 - e^{-2 \gamma t} - e^{-6\gamma t})\eta^2(\theta)],
\end{equation}
decreases with increasing $\eta$.  At $t = \tau'_{\rm d}$, ${\cal F}^{(p)}[\mu(t), |\Psi\rangle] = 1/2$ if $\eta = 1$.  Therefore, the teleportation 
fidelity may be better than $2/5$ even though there's zero entanglement left in the teleported state.  Entanglement teleportation is certainly more 
demanding \cite{Lee}.

Now, suppose Alice and Bob share the ideal channel states $\chi_0 \otimes \chi_0$ but $\alpha = d$.  We 
note that since the channel states are ideal, the states of Bob's particles obtained after Alice's measurement have the exact amount of 
entanglement as the original input state.  Bob's recovery operations are local operations that will decrease the entanglement between his particles  
if they are noisy.  Indeed, we find that other than measurement outcomes $(j_1, j_2) = (1, 1)$, $(1, 2)$, 
$(2, 1)$ and $(2, 2)$; all noisy recovery operations result in ESD of $\tilde{{\cal E}}^{(d)}_{\chi_0, mn}(|\Psi\rangle\langle\Psi|)$ at time 
$\tau'^{(d)}_{\rm d}$ (see Table I).
\begin{table}[t]
\begin{tabular}{|c|l|l|l|l|l|l|}
\hline
$(j_1, j_2)$            & (3, 3)   & (0, 3), (3, 0) & (1, 3), (3, 1) & (0, 0)   & (0, 1), (1, 0) & (0, 2), (2, 0), (2, 3), (3, 2) \\ \hline
$\tau'^{(d)}_{\rm d}$   & 4.41327  & 4.26935        & 4.82192        & 4.47320  & 4.82192        & 8.82654 \\ \hline
$\tau''^{(d)}_{\rm d}$  & 0.673553 & 0.620059       & 0.798830       & 0.636653 & 0.769228       & 0.846130 \\ \hline
\end{tabular}
\caption[Table I]{Alice's measurement outcomes are $j_1$ and $j_2$.  $\tau'^{(d)}_{\rm d}$ and $\tau''^{(d)}_{\rm d}$ are obtained assuming 
$\theta = \pi/4$, $\gamma = 1/10$, and $\omega_0 = 1$.  In addition, for $\tau''^{(d)}_{\rm d}$, we have $t_0 = 10$.}
\end{table}
For given $\theta$ and $\gamma$, $\tau'^{(d)}_{\rm d}$ depends on $\omega_0$ (see Table II).
\begin{table}[t]
\begin{tabular}{|c|l|l|l|l|l|l|l|l|l|l|l|}
\hline
$\omega_0$            & 0.5     & 0.6     & 0.7     & 0.8     & 0.9     & 1.0     & 1.1     & 1.2     & 1.3     & 1.4     & 1.5 \\ \hline
$\tau'^{(d)}_{\rm d}$ & 4.20973 & 4.14431 & 4.16111 & 4.22990 & 4.27950 & 4.26935 & 4.22880 & 4.18849 & 4.16516 & 4.16827 & 4.19592 \\ \hline
\end{tabular}
\caption[Table II]{Alice's measurement outcomes are $j_1 = 3$ and $j_2 = 0$.  $\tau'^{(d)}_{\rm d}$ is obtained assuming $\theta = \pi/4$ and 
$\gamma = 1/10$.}
\end{table}
The existence of ESD here implies that to ensure $\tilde{{\cal E}}^{(d)}_{\chi_0, mn}(|\Psi\rangle\langle\Psi|)$ has nonzero negativity and fidelity 
${\cal F}^{(d)}_{mn}[\chi_0, |\Psi\rangle] \equiv \langle\Psi|\tilde{{\cal E}}^{(d)}_{\chi_0, mn}(|\Psi\rangle\langle\Psi|)|\Psi\rangle$ larger than $2/5$ 
there is a critical $\omega_0$ below which these are impossible to achieve.  For instance, to have 
${\cal N}[\tilde{{\cal E}}^{(d)}_{\chi_0, 12}(|\Psi\rangle\langle\Psi|)] \not= 0$ and ${\cal F}^{(d)}_{12}[\chi_0, |\Psi\rangle] > 2/5$, we require that 
the minimum $\omega_0$ be between $0.5$ and $0.6$, when $\theta = \pi/4$ and $\gamma = 1/10$.  When $\chi = \zeta(t_0)$ and $\alpha = d$, 
$\tilde{{\cal E}}^{(d)}_{\zeta(t_0), mn}(|\Psi\rangle\langle\Psi|)$ suffers ESD at $\tau''^{(d)}_{\rm d}$.  In general, 
$\tau''^{(d)}_{\rm d} < \tau'^{(d)}_{\rm d}$ (see Table I).

In conclusion, we have revealed further interesting dynamical properties of quantum systems when they are subject to non-commuting noises in our study.  
These results are relevant to protocols, such as teleportation-based computation, where teleportation has to be considered as a dynamical process.


\begin{thebibliography}{99}
\bibitem{remark1} The same result holds when $|\Psi^2_{\rm Bell}\rangle$ is subject to $L_{10} = \sigma^1 \otimes \sigma^0$ and $L_{03}$.  It is also 
not necessary to subject the state to the noises simultaneously.
\bibitem{remark2} A detailed report of this study will be presented somewhere else.
\end{thebibliography}
\end{document}